\documentclass[preprint,12pt]{elsarticle}
\usepackage{amsmath}
\usepackage{graphicx}
\usepackage{booktabs,subfigure}
\usepackage{array}
\usepackage{gensymb}
\usepackage{color}
\usepackage{subfig}
\usepackage{caption}
\usepackage[svgnames]{xcolor}
\begin{document}

\journal{Physica B: Condensed Matter}

\begin{frontmatter}

\title{First principles investigation of zb-TiSn: A promising narrow bandgap semiconductor}
\cortext[cor1]{Corresponding author}
\author[1]{Sudeep R}
\author[1]{Sarojini M}
\author[1]{Uma Mahendra Kumar Koppolu\corref{cor1}}%
\ead{umamahendra@vit.ac.in}
\affiliation[1]{addressline={Dept of Physics, School of Advanced Sciences},
             organization={Vellore Institute of Technology},
             city={Vellore},
             postcode={632014},
             state={TN},
             country={Bharat(India)}}
	
\begin{abstract}
 We have investigated the structural stability of a binary compound TiSn in the zincblende symmetry. The phonon dispersion studies confirms that, TiSn with a nominal composition of 1:1 can exist in zincblende form. No imaginary frequencies are observed indicating the stable bonding nature of Ti-Sn. From the First principles calculations based on density functional theory, the resulting electronic band structure had revealed that zb-TiSn, is a narrow band gap semiconductor with an energy gap of 0.3 eV with GGA-PBE. The bonding nature is identified as polar covalent, determined from charge density difference plots and Bader charge analysis. Further more, the linear optical properties of zb-TiSn are derived from the Khon-Sham eigenvalues.
\end{abstract}

\begin{graphicalabstract}
\includegraphics[scale=0.5]{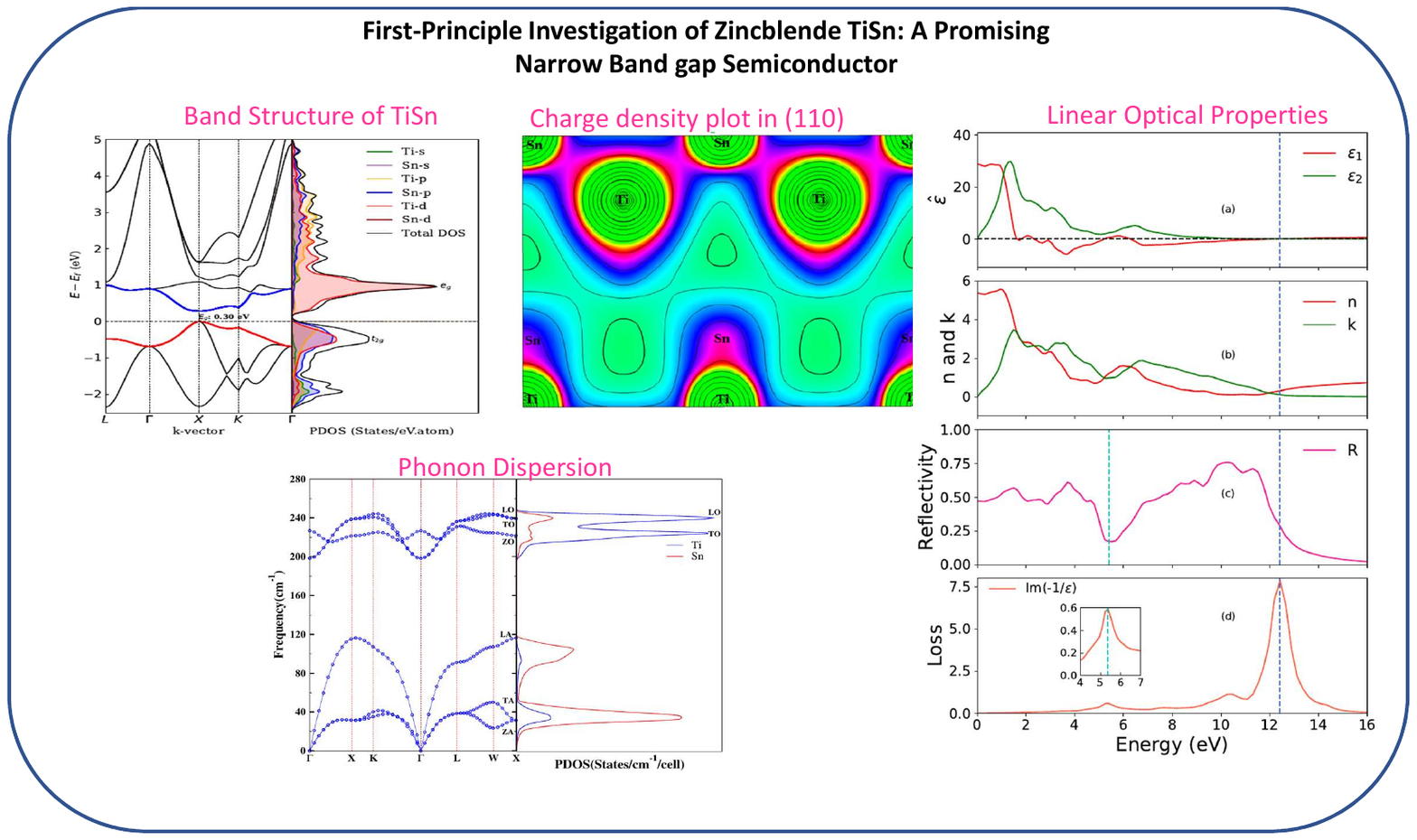}
\end{graphicalabstract}

\begin{highlights}
\item TiSn in zincblende form is structurally and dynamically stable.  
\item Phonon dispersion shows no imaginary frequencies.  
\item DFT calculations confirm a narrow band gap of 0.3 eV (GGA-PBE).  
\item zb-TiSn is identified as a narrow-gap semiconductor.  
\item Bonding is polar covalent from charge and Bader analysis.  
\item Optical properties suggests that zb-TiSn is applicable in IR-photonics.
\end{highlights}

\end{frontmatter}
\section{Introduction}
	Narrow bandgap semiconductors with energy gap ranging between 0.2 and 0.3 eV, such as Pb-chalcogenides, In$_{1-x}$Ga$_x$As, In$_{1-x}$Ga$_x$Sb, and Si$_{1-x}$Ge$_x$, are particularly effective for thermoelectric power generation through the Seebeck effect, efficient at low temperatures\cite{black}. Bismuth telluride (0.14 eV), when alloyed with Sb$_2$Te$_3$ (0.24 eV), exhibits high thermoelectric efficiency near room temperature due to its combination of high carrier mobility and low thermal conductivity, making it suitable for Peltier cooling and waste heat recovery\cite{thermoelectricbismuth}. PbTe (0.32 eV) and its alloys are among the most advanced thermoelectric materials for intermediate temperatures\cite{pbte}, broadening their practical applications. Additionally, Bi$_2$Te$_3$ and Sb$_2$Te$_3$ have demonstrated high sensitivity in temperature sensors, often surpassing conventional thermocouples in microelectronic applications\cite{microelectronic}.
	
Beyond thermoelectrics, narrow bandgap materials ($E_{g} < 0.5$ eV) plays a crucial role in photovoltaics (PV) by efficiently absorbing the low-energy photons and enhances the output current density and enables the waste heat utilization in thermophotovoltaic systems. These compound are further suitable in laser power beaming under mid-infrared (3-5 $\mu$m) light which supports the remote energy delivery and waste heat recovery\cite{photovoltaic}. 

In addition to energy-related applications, narrow band gap semiconductors, like InSb (band gap $\sim$0.23 eV), exhibits strong thermal excitation of intrinsic carriers, leading to high carrier concentrations. Unlike wide band gap materials, they face doping challenges and have softer mechanical properties, shaping their infrared optoelectronic behavior\cite{doping}. Their ability to generate non-equilibrium electron–hole pairs under radiation alters the electrical transport and enables the high-sensitivity infrared detection. With small effective masses, high electron mobility, long carrier lifetimes, strong optical absorption, and low thermal generation, they ensure fast response in infrared sensing and communication\cite{physics}. Additionally, the tunable band gaps further enhance their versatility across the infrared spectrum\cite{inassb}. 

Similarly, semiconductors such as HgCdTe, with a tunable band gap ranging from approximately 0.14 eV to 0.35 eV, have become essential in infrared detector technology\cite{hgznte}. HgCdTe is a pseudo-binary alloy semiconductor that crystallizes in a zincblende structure and due to its band gap tunability over the 1-30 $\mu m$ range, it has evolved into the most versatile material for infrared detection\cite{hgcdte2}, enabling advanced multispectral and dual-band imaging for space borne remote sensing applications\cite{hgcdte3}. The small band gaps of these materials also make them ideal for ultrafast photonics. PbS (0.4 eV) and Bi (0.14 eV) are particularly advantageous for mode-locked pulse lasers, as narrower band gaps enable broader light absorption, enhancing mode-locking performance and expanding their applications in optical communication and laser technologies\cite{optonarrow}. Hence, there is considerable interest in designing and developing narrow band gap semiconductors for various applications. 

The experimentally obtained Sn-Ti phase diagram reveals the presence of five intermetallic compounds; SnTi$_3$\cite{snti3}, SnTi$_2$\cite{snti2}, Sn$_3$Ti$_5$\cite{sn3ti5}, Sn$_5$Ti$_6$\cite{sn5ti6} and Sn$_3$Ti$_2$. Among them, the existence of Sn$_3$Ti$_2$ was first reported by Kuper et al\cite{sn3ti2}. Interestingly, Sn$_5$Ti$_6$ occurs in both hexagonal and orthorhombic forms, with the orthorhombic structure identified through crystallization experiments of Sn-rich Sn–Ti melts\cite{orthorhombic}.

First-principles calculations by Catherine Colinet et al.,\cite{structural} have been used to determine the `\textit{enthalpy of formation}' of Sn–Ti intermetallic compounds through an \textit{ab-initio} approach. The predicted ground-state structures are consistent with those stable at low temperatures. The calculations confirm that hexagonal Sn$_5$Ti$_6$ (hexa-Sn$_5$Ti$_6$) is more stable than its orthorhombic counterpart (ortho-Sn$_5$Ti$_6$), establishing that the orthorhombic phase is not a low-temperature ground state. These insights contribute to understanding the Sn–Ti system`s ground-state and metastable phases and have been integrated into thermodynamic databases for CALPHAD modeling. \textcolor{blue}{As well in the open quantum material database(OQMD),several of Sn-Ti intermetallic compounds are explored using DFT. The zincblende structured TiSn is reported as metastable\cite{oqmd}. However, our lattice dynamic studies have predicted otherwise. Further, in the manuscript the phonon properties of zb-TiSn might shed some light on the stability of the compound. It is suggested that, in future the metastable phases could be experimentally realised in various ways, like epitaxial growth on suitable substrates, which is proven to be successful in the case of zb-CrAs resulting in half-metallic state\cite{epi_cras}. Stabilizing the zinc blende (ZB) structure involves a combination of doping, substrate selection\cite {chic}, multilayer engineering\cite{li}, and nano-structuring \cite{kra,ding,kumar}. While the ZB phase is metastable under ambient conditions, these strategies can promote its formation and stability.} 

Previously, we have investigated TiGe in zincblende symmetry, which is also a narrow band gap semiconductor\cite{tige}. We found that TiSn also have similar properties in zincblende symmetry. In the current manuscript, we propose zincblende TiSn as a promising candidate, with a band gap of 0.2 to 0.3 eV. A detailed structural, electronic and optical characterization of zb-TiSn is presented based on careful first principle investigation. The electronic band structure is determined using the Grid-based Projector-Augmented Wave method implemented in GPAW code \cite{gpaw}. The robustness of the semiconducting nature is explored by using different exchange and correlation functional \textit{viz.}Local Density Approximation (LDA-PZ)\cite{lda}, Generalized Gradient Approximation (GGA-PBE)\cite{gga}, and Revised Perdew-Burke-Ernzerhof (GGA-RPBE)\cite{rpbe}.  A pseudo-potential plane wave code QUANTUM ESPRESSO is used to calculate the lattice dynamics and established the phase stability of the compound. The rest of the paper is structured as follows. The computational methods employed in this work are described, followed by a discussion of the electronic structure, charge distribution, Bader charge analysis results, and effective mass of charge carriers is detailed. Later, the lattice dynamics are explored to highlight the structural stability of the material, and the optical properties are examined to determined the applicability in opto-electronics in the infra-red range.

\section{Computational methods}
	The ground state properties of zb-TiSn are determined using first-principles total energy calculations based on density functional theory (DFT) performed with the GPAW code\cite{gpaw}. The zb-TiSn model was constructed using the Atomic Simulation Environment (ASE)\cite{asr} with fundamental structural parameters obtained from the Crystallographic Open Database\cite{crystallography} (entry no. 9008845)of GaAs. The crystallographic symmetry is represented by the space group symmetry of \texttt{F-43m} (no. 216), with constituent atoms, Ti located at the Wyckoff position (0.0, 0.0, 0.0) and Sn at (0.25, 0.25, 0.25). The plane wave expansion energy cut-off is 450 eV. The Brillouin zone (BZ) is sampled using an evenly distributed grid of $(4\times4\times4)$ k-points mesh. In order to establish the robustness of the compound's semiconducting nature , we have implemented three different exchange and correlation functions, namely LDA-PZ, GGA-PBE, and GGA-RPBE, and examined the effect of functionals on the structural parameters and electronic band structure of zb-TiSn. \textcolor{blue}{In order to derive the band gap energy accurately, we have used two methods, one is GLLB-sc fuctional which is known to predict the band gap energy accurately and the second method is GGA-PBE+U, where we have applied the Hubbard U correction to the Ti-d states to compensate the Coulombic repulsion in the 3d-transition metals.} 
	
	To determine the phase stability of zb-TiSn, complete phonon dispersion curves were computed at high-symmetry points following the k-path ($\Gamma$-X-K-$\Gamma$-L-W-X) in the irreducible Brillouin zone (BZ). The phonon density of states for zb-TiSn are also calculated along with the phonon dispersion curves. The calculations were performed within the framework of Density Functional Theory (DFT), employing both the Generalized Gradient Approximation (GGA-PBE)\cite{gga} and the Local Density Approximation (LDA-PZ)\cite{lda}.
The following pseudo-potentials are used for phonon calculations taken from the \texttt{pslibrary.1.0.0}. For LDA-PZ functional, \texttt{Ti.pz-spn-rrkjus-psl.1.0.0.UPF}, \texttt{Sn.pz-dn-rrkjus-psl.1.0.0.UPF }and for GGA-PBE functionals we have used, \texttt{Ti.pbe-dn-rrkjus-psl.1.0.0.UPF}, \texttt{Sn.pbe-dn-rrkjus-psl.1.0.0.UPF} are used. 
The Monkhorst-Pack $k$-point sampling method was used to integrate across the first Brillouin zone with a $k$-point set of $8 \times 8 \times 8$ and a plane wave cutoff energy of 60 Ry. To ensure convergence of phonon frequencies and free energy, phonon dynamical matrices were generated using a $2 \times 2 \times 2$ $q$-grid in the irreducible wedge of the first BZ at high-symmetry directions, characterized by both LDA-PZ and GGA-PBE approximations. The phonon calculation work-flow is automated using \texttt{Thermo-PW}, a Fortran wrapper for QUANTUM ESPRESSO package.

\section{Results and Discussion}
\subsection{Structural optimization}

The ground state structural parameters are optimised by following the conventional method of curve fitting the total energy vs volume data with equation of state(EOS)\cite{murnaghan}. The energy-volume relationship from the Murnaghan \textit{`equation of state'} provides more accurate description of a material's behavior under different conditions (e.g., varying volume, pressure). The equilibrium lattice constant, bulk modulus, and equilibrium volume are tabulated in Table \ref{tab_struc} for different DFT functionals.
\begin{equation}
E(V) = E_0 + \frac{B_0 V}{B_0' } \left( \frac{(V_0/V)^{B_0'}}{B_0' -1} +1 \right) - \frac{B_0 V_0}{B_0' -1}
\end{equation}

where,$V_0$ is the Equilibrium volume, $V$ is the current volume,$B_0$ is the bulk modulus at equilibrium volume and $B_0'$ is the derivative of the bulk modulus with respect to volume.

	\begin{table}[h]
		\centering
		\begin{tabular}{lllll}
			\toprule
			\textbf{Structural Parameters} & \textbf{LDA-PZ} & \textbf{GGA-PBE} & \textbf{GGA-RPBE} & \\ 
			\midrule
			Lattice constant $a$ (\AA{}) &6.15  &6.33  &6.40   \\
			Volume $V_{0}$ (\AA{}$^3$) &232.18  &252.82  &262.09  &  \\
			Bulk modulus $B_0$(GPa) &54.0  &36.37  &40.86   \\
			 $B_0'$&4.444  &3.934  &3.756   \\
			Bond length (Ti-Sn) (\AA{}) &2.663  &2.741  &2.771   \\
			\bottomrule
		\end{tabular}
		\caption{Structural parameters calculated using different exchange-correlation functionals.}
		\label{tab_struc}
	\end{table}
	
\textcolor{blue}{The optimized lattice parameter of zb-TiSn is a = 6.15 \AA{} with LDA-PZ, 6.33 \AA{} with GGA-PBE, and 6.40 \AA{} with GGA-RPBE.} 
As it is well known that LDA-PZ will generally result in shorter bond lengths and smaller lattice parameters compared to GGA-PBE and other flavours of GGA. The same has been observed in the current studies. The estimation of Bulk modulus $B_{0}$ also follows a similar trend. Because of the shorter bond lengths in LDA-PZ, the Bulk modulus$B_{0}$ has reached a value of 54 GPa, but decreases to 36.37 GPa with GGA-PBE and 40.86 GPa with GGA-RPBE. \textcolor{blue}{In summary, the bulk modulus $B_{0}$ values suggest that zb-TiSn is a relatively
soft material}.Currently, no experimental results are available to compare with.
	
\subsection{Bandstructure of TiSn}

\begin{figure}
	\centering
\subfigure[LDA-PZ]{\includegraphics[scale=0.38]{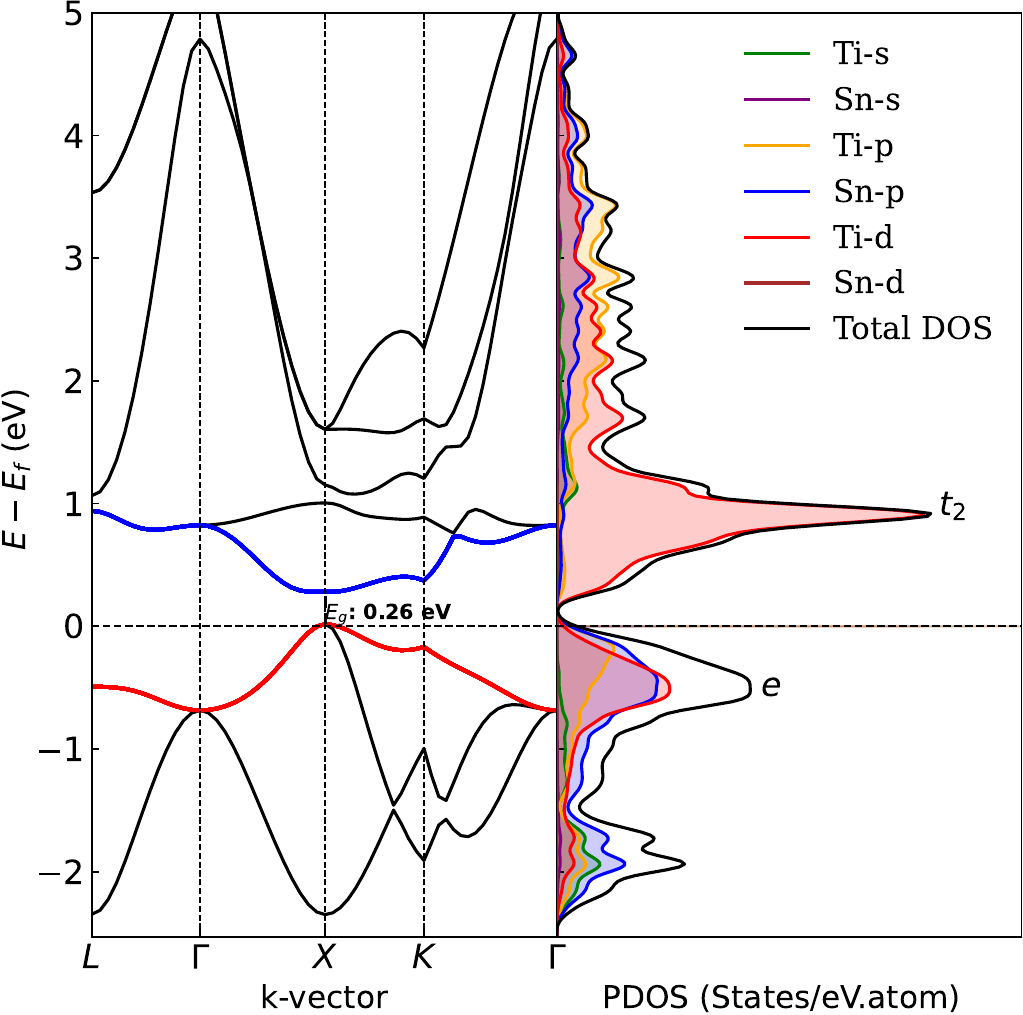}}
\subfigure[GGA-PBE]{\includegraphics[scale=0.38]{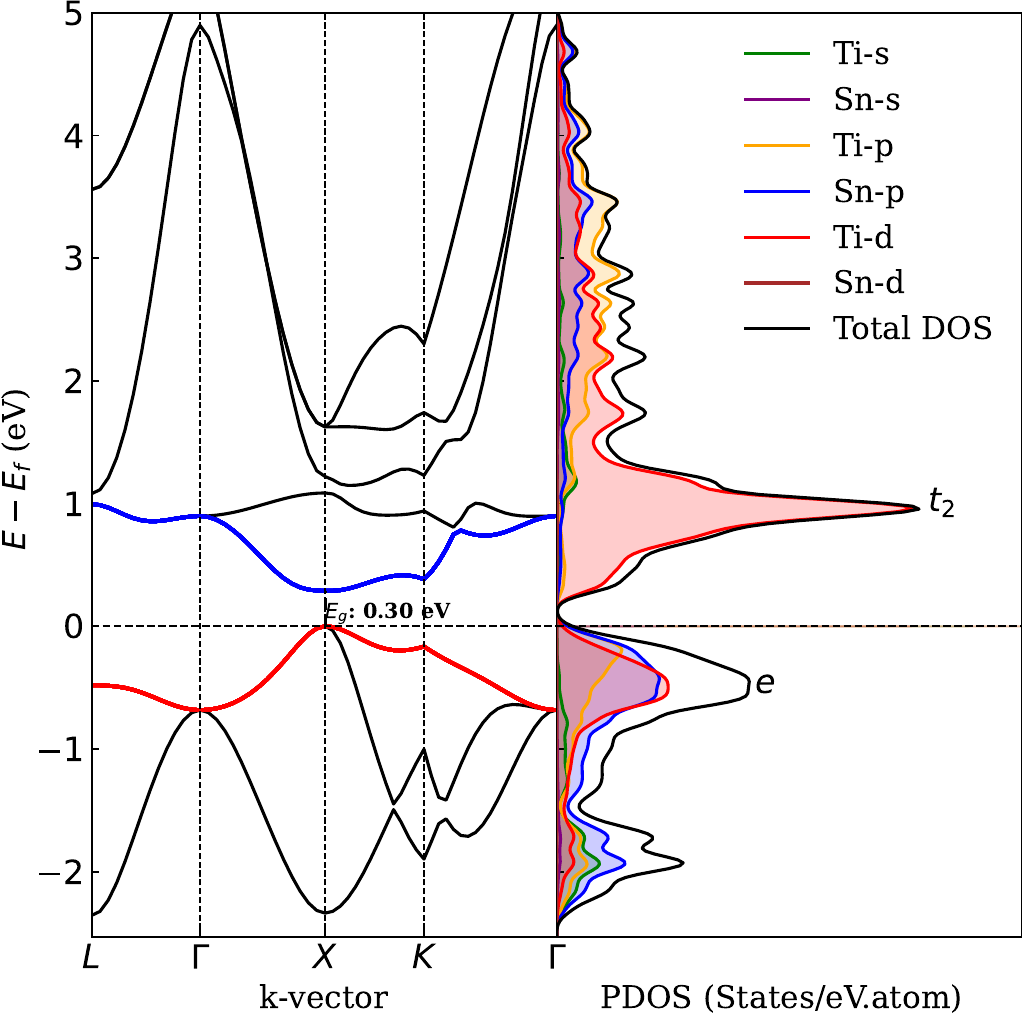}}	
\subfigure[GGA-RPBE]{\includegraphics[scale=0.38]{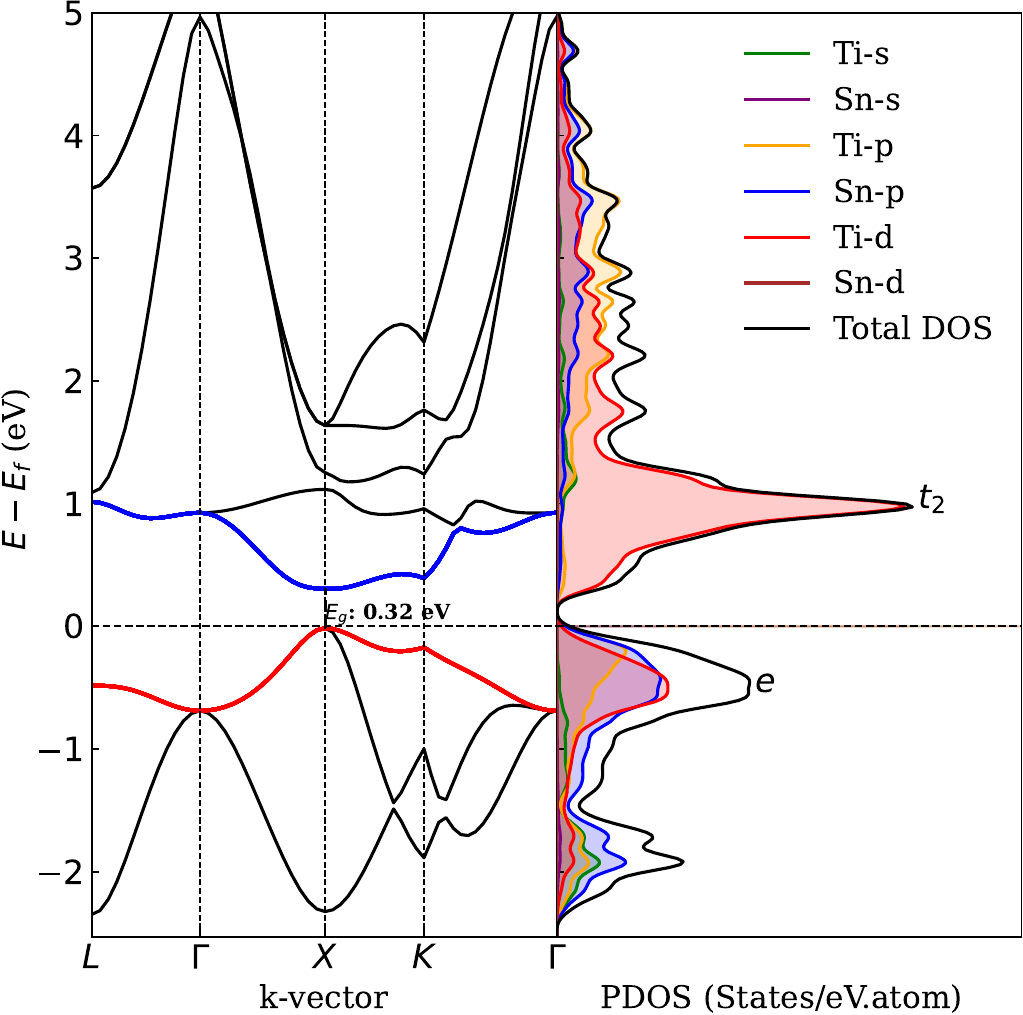}}
\subfigure[GLLB-sc]{\includegraphics[scale=0.38]{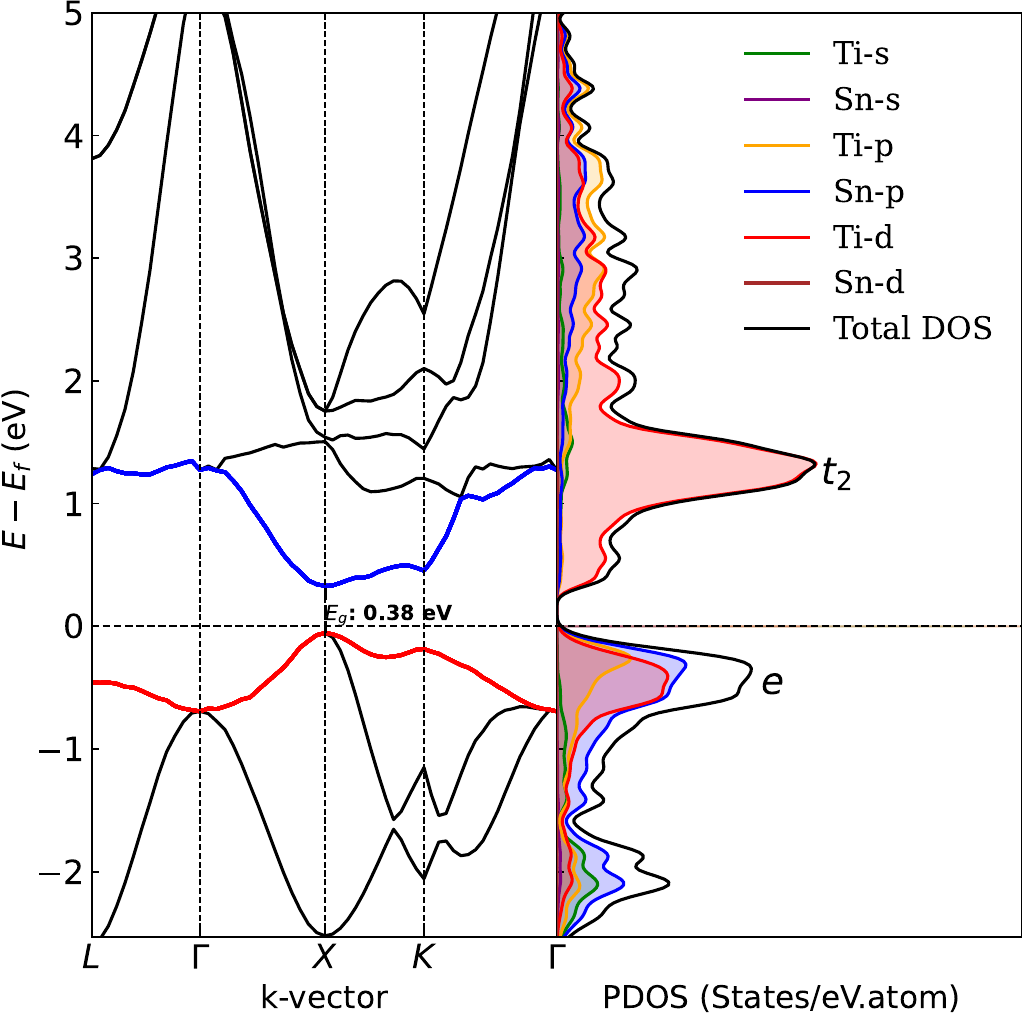}}
	\caption{Electronic band structure and PDOS (Projected density of states) of zb-TiSn with different exchange-correlation functionals (a) LDA-PZ, (b) GGA-PBE, (c) GGA-RPBE and d)GLLB-sc. After the band splitting the degenerate orbitals are marked as $e$ and  $t_{2}$.}
	\label{bs}
\end{figure}

Along the high symmetry points in the Brillioun zone (L-$\Gamma$-X-K-$\Gamma$), the electronic band structure is calculated using three different exchange and correlation functionals, namely Local Density Approximation (LDA-PZ), Perdew-Burke-Ernzerhof (GGA-PBE), and Revised Perdew-Burke-Ernzerhof (GGA-RPBE)and the resulting band structure is shown in the Figure \ref{bs} a,b and c.

The ground state electronic band structure of zb-TiSn exhibits a direct and narrow band gap at high symmetry point `X' in the Brillioun zone. \textcolor{blue}{This direct band gap effectively supports the transition of optically generated electrons, greatly increasing the rate of electronic transitions\cite{directband}. To investigate the contributions of atomic states near the Fermi surface, the energy axis is set from -2.5 eV to +5 eV. The projected density of states (PDOS) reveals that the Ti d-states, Ti p-states, and Sn p-states contribute significantly in this region. This indicates that electrons from these Ti d-state, Ti p-state and Sn p-state are more likely to transition to the conduction band minimum and play a crucial role in determining the electronic properties of zinc blende Ti-Sn \cite{fermi-surface}.}

\textcolor{blue}{It is evident from the projected density of states that the valence band is comprised of Ti-3d states and Sn-5p states, resulting from of p-d hybridization. 
In tetrahedral coordination of zb-TiSn, the five $d$-orbitals of Ti will split into two sets due to the electrostatic field from four Sn-ligands positioned at the corners of a tetrahedron. The $t_2$ orbitals $(d_{xy}, d_{yz}, d_{xz})$ experience more repulsion as they point closer to the ligand directions, making them higher in energy, while the $e$ orbitals $(d_{z^2}, d_{x^2 - y^2})$ point between ligands and are lower in energy. Since the $e$ orbitals are lower in energy in tetrahedral coordination, electrons fill them first according to the Aufbau principle. These occupied low-energy states contribute to the valence band. In contrast, the $t_2$ orbitals are higher in energy and partially occupied, contributing mainly to the conduction band as available excited states.}

\begin{table}[]
	\centering
	\begin{tabular}{l c}
\toprule
$f_{xc}$ & Band gap(eV) \\
\midrule
LDA-PZ   &0.26 eV  \\
GGA-PBE   &0.30 eV \\
GGA-RPBE  &0.32 eV  \\
GLLB-sc   & 0.38 eV \\
\bottomrule
	\end{tabular}
	\caption{Bandgap energy of zb-TiSn at high symmetry point `X' for different exchange and correlation functionals}
	\label{tab:bandgap}
\end{table}

Based on the nature of the exchange and correlation functionals, the estimated band gap is listed in Table \ref{tab:bandgap}. zb-TiSn exhibits relatively small band gap of 0.26 eV with LDA-PZ, 0.30 eV with GGA-PBE and 0.32 eV with GGA-RPBE functionals.\textcolor{blue}{To assess the robustness of the electronic structure, calculations were performed using the GLLB-sc functional, which incorporates a derivative discontinuity correction which is absent in ordinary GGA and LDA functionals. The converged lattice constant was found to be 6.215 Å, and the calculated band gap was 0.38 eV. This band gap is in qualitative agreement with band gap values obtained from LDA-PZ, GGA-PBE, and GGA-RPBE functionals and supports that fact that zb-TiSn could be a narrow band gap semiconductor. Further, we have used GGA-PBE+U method to study the effect of Hubbard `U' correction on the lattice parameter and the resulting band gap. However, the results indicated that the band gap appeared to be very similar across the entire U range, suggesting that the effect of Hubbard U correction on the band gap energy is marginal in zb-TiSn. }


\subsection{Effective mass}
Effective mass is an important material parameter that governs most of the carrier transport properties in semiconductors. Precise knowledge of effective mass is important for the design technology of optoelectronic devices\cite{broad}. Since the effective mass directly influences the mobility of charge carriers, it plays a significant role in determining the performance and efficiency of these devices. 

The effective mass is calculated using the expression\cite{kittel}
\begin{equation}
	\frac{1}{m^*} = \frac{1}{\hbar^2} \frac{d^2E}{dk^2}
\end{equation}

By fitting the actual $E-k$ dispersion curve around conduction band minima (CBM) and valance band maxima (VBM) of energy eigenvalues, the effective mass of electrons and holes can be obtained. We have used the $4th$-order polynomial fit as the band extrema in zb-TiSn are appearing not strictly parabolic and hence we have used the following expression to fit the $E-k$ dispersion curve.

\begin{equation}
	E(k) = \alpha k^4 + \beta k^3 + \gamma k^2 + \delta k + \epsilon
	\label{eff}
\end{equation}
The first and second-order differentiation of equation \ref{eff}  with respect to momentum space gives
\begin{equation}
	\frac{dE}{dk} = 4\alpha k^3 + 3\beta k^2 + 2\gamma k + \delta
	\label{first}
\end{equation}
 and 
\begin{equation}
	\frac{d^2E}{dk^2} = 12\alpha k^2 + 6\beta k + 2\gamma
	\label{sec}
\end{equation}

Thus, the effective mass can be obtained as
\begin{equation}
	m^* = \frac{\hbar^2}{12\alpha k^2 + 6\beta k + 2\gamma}
\end{equation}
Where $m^*$ is the effective mass and $\alpha$, $\beta$, $\gamma$ and $\delta$ are the coefficients of the polynomials used for curve fitting. 

Since the effective mass is related to the curvature of the band extrema, setting the first derivative of E(k) to zero and solving the resulting cubic polynomial equation identifies the extrema (critical points) of the electronic band structure.  The effective mass of electrons and holes for TiSn are calculated by fitting the preuso-parabolic curves at CBM and VBM in the band structure of TiSn located at high symmetry point `X’ of the Brillouin zone. With different exchange correlation functions, namely LDA-PZ, GGA-PBE, and GGA-RPBE, the calculated effective mass are tabulated in Table \ref{effective_mass}.

\begin{table}[h]
		\centering
		\renewcommand{\arraystretch}{1.5}
		\begin{tabular}{lcc}
			\toprule
			\textbf{$f_{xc}$}   & \textbf{Conduction Band }   & \textbf{Valence Band } \\
			                      & ($\Gamma$-X-K)              & ($\Gamma$-X-K) \\
			\midrule
			LDA-PZ  &  \( 1.80 \, m_0 \) &  \( -0.63 \, m_0 \) \\
			GGA-PBE  &  \( 2.88 \, m_0 \) &  \( -0.77 \, m_0 \) \\
			GGA-RPBE &  \( 3.60 \, m_0 \) &  \( -0.97 \, m_0 \) \\
			\bottomrule
		\end{tabular}
		\caption{Effective Mass of the charge carriers near the conduction band minima and valance band maxima are listed.}
		\label{effective_mass}
	\end{table}

At the high-symmetry point `X', the effective mass of charge carriers at the CBM is significantly larger than that at the VBM due to the flatness of the CBM. The lower effective mass at the VBM facilitates the excitation of charge carriers to the CBM. However, the higher effective mass at the CBM restricts their mobility, causing them to remain localized. This localization of charge carriers at the CBM results in minimal dispersion, making electrons appear much heavier. This significantly reduces their kinetic energy, which causes electron interactions to dominate and giving rise to interesting physical phenomenon such as fractional quantum Hall effect\cite{high, nearly}, Wigner crystallization\cite{Wigner}, Bose-Einstein condensation\cite{bose}, and high-temperature superconductivity\cite{superconductivity, superfluidity}.

\subsection{Charge density distribution}

In DFT, the electron density $\rho(r)$ describes the distribution of electrons in a material, providing insights into bonding characteristics. Using GPAW, the charge density is obtained by solving the self-consistent Kohn-Sham equations\cite{kohn}. The charge density is then reconstructed from the occupied Kohn-Sham orbitals using the equation \ref{rho},
\begin{equation}
	\rho(r) = \sum_i f_i |\psi_i(r)|^2
	\label{rho}
\end{equation}
Where, $\psi_i(r)$ is Kohn-Sham wave function, $f_i$ is occupation number from Fermi-Dirac distribution and $|\psi_i(r)|^2$ represents the probability density. 

The nature of the chemical bond between two atoms can be determined by the difference in their electronegativity values. According to the Pauling scale, the electronegativity values for Ti is 1.54\cite{ti} and Sn it is 1.91\cite{sn} and their electronegativity difference is $(\Delta\chi) = 0.42$ for the Ti-Sn bond. When the electronegativity difference is $(\Delta\chi < 1.7)$, that would indicate covalent nature of the bond  \cite{electronegativity}. With 0.42 as the electronegativity difference, Ti-Sn bond appears to be covalent as all other zincblende structured compounds. 

\begin{figure}
	\centering
\subfigure[]{\includegraphics[scale=0.35]{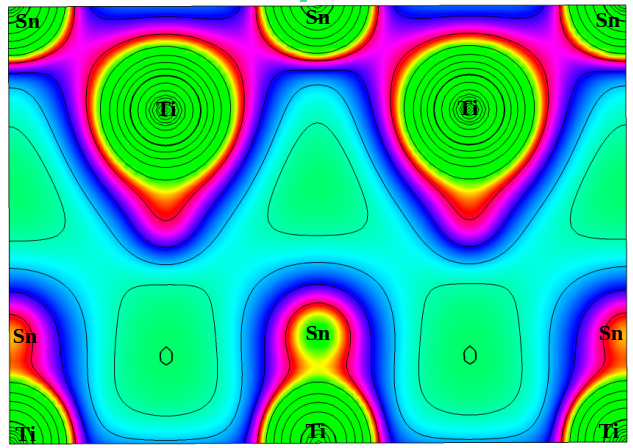}}
\subfigure[]{\includegraphics[scale=0.35]{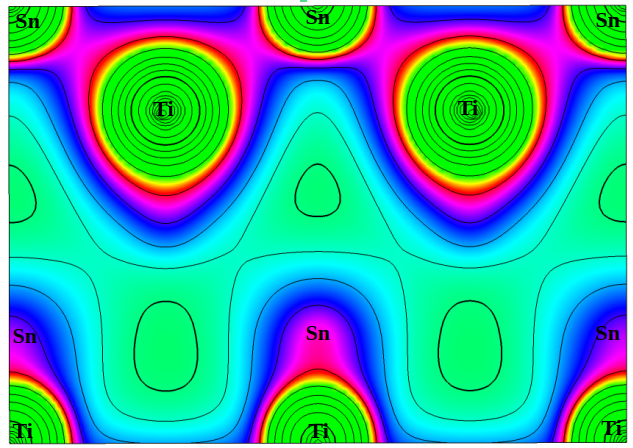}}
\subfigure[]{\includegraphics[scale=0.35]{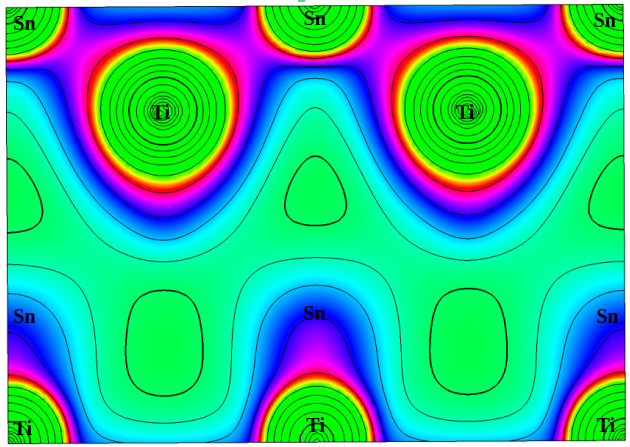}}
	\caption{The electron density plot along (1 1 0) crystallographic plane of TiSn under a)LDA-PZ b)GGA-PBE c)GGA-RPBE }
	\label{charge}
\end{figure}

Along the (1 1 0) crystallographic plane, with projection range of LDA-PZ(3.8-6 e/Å$^3$), GGA-PBE(3.8-6.2 e/Å$^3$), and GGA-RPBE(3.8-6.3 e/Å$^3$), the electron density plots of zb-TiSn are shown in Figure\ref{charge} a,b and c. The extended charge distribution of the denser electron cloud from Ti to Sn suggests covalent bonding, which results from p-d hybridization, as evidenced in the PDOS shown in Figure \ref{bs}. Observing the charge density plots reveals that the bonds between Ti and Sn is predominantly covalent and with certain degree of ionic character. 

From Figure \ref{charge} a, b and c, it is evident that, the exchange and correlation functionals has distinct effect on charge distribution. In the Figure \ref{charge}a, we show the charge density plot calculated with LDA-PZ. The contours are more tightly packed around the atoms Ti and Sn suggesting stronger localization of electrons. The bonding regions (between atoms) show less charge delocalization. With GGA-PBE in Figure \ref{charge}b, the charge density is more smeared out, particularly in the bonding regions. The contours are slightly more diffuse compared to LDA-PZ, indicating reduced localization. Additionally, the outer contours are exhibiting a small distortion from their regular circular shape and extend into interstitial regions (seen more in GGA-PBE), suggesting that the charge is spreading due to hybridization between Ti and Sn atoms. 

Based on the color code we could interpret the charge density plots as follows. Green representing the high charge density which is evident around atomic cores of Ti and Sn, indicating localized electron density. Red/Purple is representing the intermediate or medium density signifying the bonding regions where charge accumulates. Blue/Cyan represents the low charge density regions indicating the charge depletion with low electron density. The interstitial regions, where charges will get depleted and no charge contours can be observed. In LDA-PZ as shown in  Figure\ref{charge} a, no charge contours because of higher localization and in GGA-RPBE (Figure\ref{charge} c), the charge density has been depicted as slowly varying between the atoms signifying the covalent nature of the bond. However, GGA-PBE suggests that the bond can be polar covalent. This discrepancy can be clarified by quantifying the amount of charge which has been shared or transferred using Bader charge analysis.

\subsection{Bader Charge Analysis}
In a crystal, electrons are shared between atoms, making it difficult to define atomic charge and the amount of charge transfer.
The Bader method\cite{fast}\cite{improved}\cite{grid} solves this by partitioning the electron density $\rho(r)$ of the total space into separate basins (regions), where each basin corresponds to an individual atom, called atomic basin.

Comparing the total number of electrons in the atomic basin of an atom with its neutral atom reveals the amount of charge transfer.
Using Bader charge analysis, the amount of charge transfer between Ti and Sn in the zincblende structure under LDA-PZ, GGA-PBE, and GGA-RPBE functionals is computed and tabulated in Table~\ref{tab:charge_transfer}.

\begin{table}[h]
	\centering
	\begin{tabular}{ccc}
		\hline
		Functional & Titanium (Ti) Charge (e) & Tin (Sn) Charge (e) \\
		\hline
		LDA-PZ  & -1.05  & 1.05  \\
		GGA-PBE  & -1.11   & 1.11   \\
		GGA-RPBE & -1.14   & 1.14   \\
		\hline
	\end{tabular}
	\caption{Bader charge transfer analysis of TiSn using different functionals.The -ve sign indicates the loss of charge.}
	\label{tab:charge_transfer}
\end{table}

Bader charge analysis sheds light on zb-TiSn bonding behavior, showing that Ti loses $-1.05e$ and Sn gains $1.05e$ in LDA-PZ, while in GGA-PBE, Ti loses $-1.11e$ and Sn gains $1.11e$ and the GGA-RPBE values are similar for Ti and Sn as shown in the table \ref{tab:charge_transfer}. Charge transfer in TiSn creates a partial negative charge on Sn and a partial positive charge on Ti, confirming the polar covalent nature of the Ti-Sn bond. 

\section{Lattice dynamics of zb-TiSn} 

Since the compound zb-TiSn has not been realised experimentally, we aim to establish the compound's phase stability via lattice dynamics. We have calculated the phonon dispersion along the high symmetry points spanning the entire Brillioun zone. Using DFPT method, we have computed the dynamic matrices and extracted the inter atomic force constants.

Zincblende TiSn belongs to the point group T$_{d}$(43m) and has two atoms per primitive cell, yielding six phonon branches; one longitudinal acoustic (LA), two transverse acoustic (TA and ZA), one longitudinal optic (LO), and two transverse optic (TO and ZO) modes. The optical phonons exhibits $T_2 = \Gamma_{15}$ symmetry, consisting of one LO mode and two doubly degenerate TO modes. Around the $\Gamma$ point, the transverse acoustic (TA) and longitudinal acoustic (LA) modes exhibit linear behavior, providing insights into the velocity of sound and thermal conductivity\cite{ashcroft} of the material. 

\begin{figure}
\centering
\subfigure[LDA-PZ]{\includegraphics[scale=0.5]{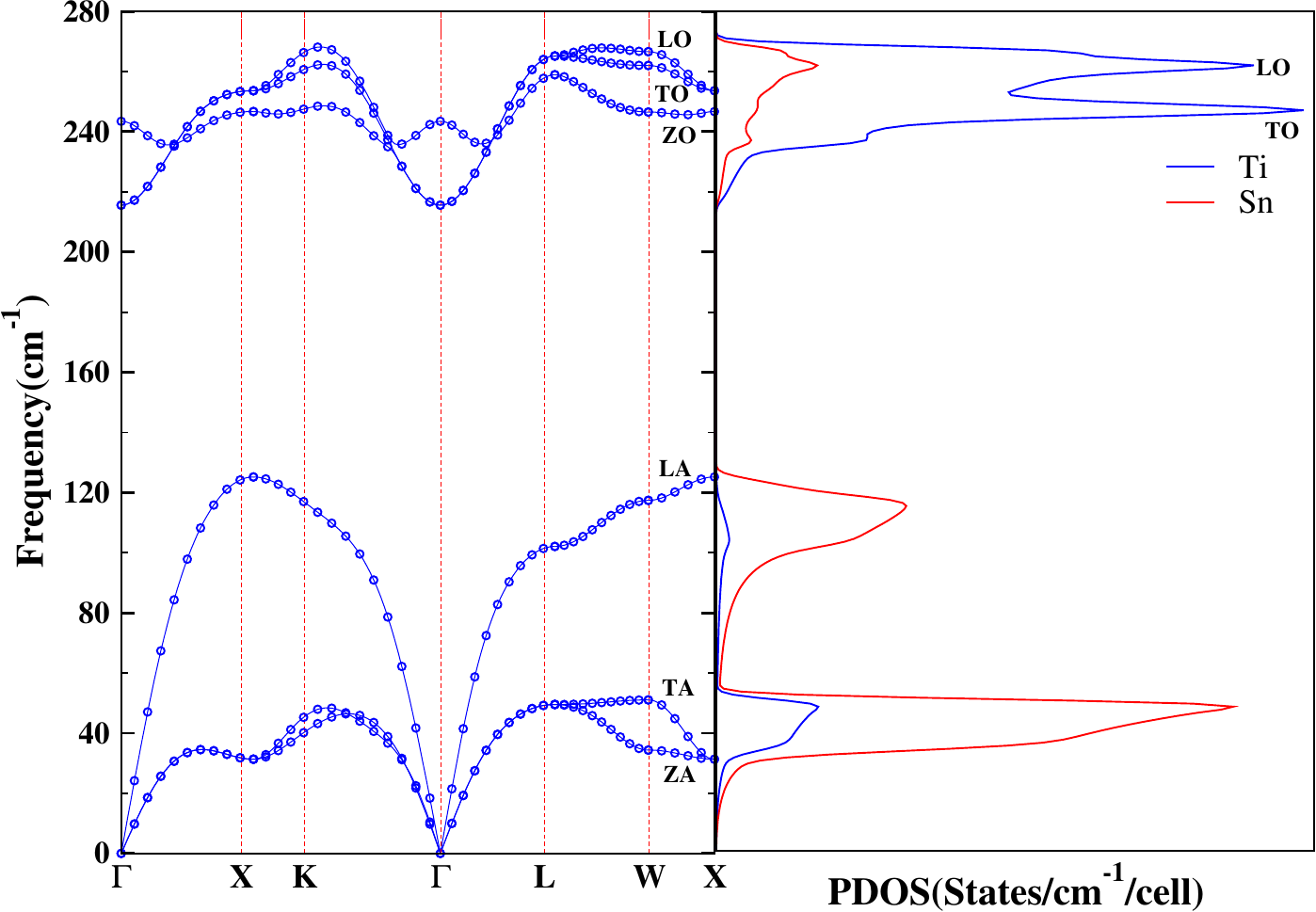}}
\subfigure[GGA-PBE]{\includegraphics[scale=0.5]{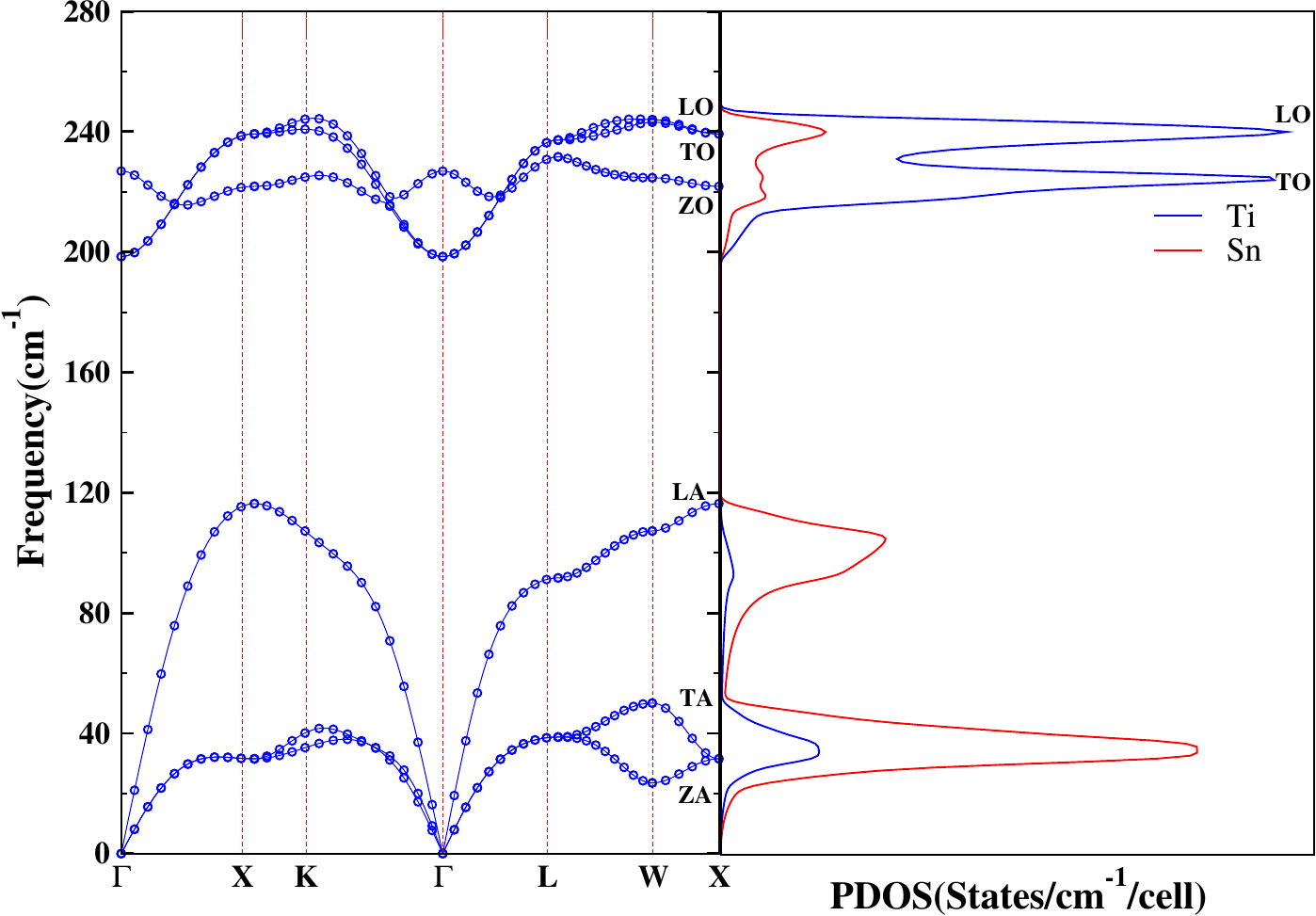}}
   \caption{Phonon dispersion and PDOS of zb-TiSn with LDA-PZ and GGA-PBE exchange and correlation functionals}
   \label{phonon}
\end{figure}

The degeneracy of the optical modes vanishes in both LDA-PZ and GGA-PBE at the $\Gamma$ point, resulting in LO-TO splitting attributed to the Lyddane–Sachs–Teller (LST) phenomenon\cite{lyddane}, as shown in Figure \ref{phonon} a and b. \textcolor{blue}{The LO-TO splitting arises from long-range Coulomb interactions in polar materials, leading to different vibrational frequencies for longitudinal and transverse optical modes at the $\Gamma$ point}. The LO modes exhibits higher energy than the TO modes due to partial ionic bonding arising from long-range Coulomb interactions, which can be described using Born effective charge tensors\cite{borneffective}. 

The Born effective charge tensors $Z_{Ti}^{*}$ and $Z_{Sn}^{*}$ were found to be diagonal, with only a single distinct component due to high symmetry. In GGA-PBE, the Born effective charges (in the units of elementary charge) are 3.98 for Ti and -3.98 for Sn, whereas in LDA-PZ, they are slightly higher at 4.56 for Ti and -4.51 for Sn. These values, which are significantly larger than the nominal ionic charges, indicate strong charge transfer and polarization, leading to a stronger response of Ti and Sn atoms to external electric fields. The deviation of the Born effective charges from their nominal values highlights the mixed ionic-covalent nature of TiSn, where Ti donates charge while Sn accepts it. 
LDA-PZ predicts higher Born effective charges due to its smaller lattice constant, which strengthens atomic interactions and enhances charge transfer. Conversely, GGA-PBE, with its larger lattice constant, weakens these interactions, leading to lower Born effective charges and reduced charge polarization.

LO-TO splitting arises from the ionic component in polar and covalent bonds. It is evident in the charge density difference plot of zb-TiSn, which reveals that the charge accumulation zones on Sn atoms and depletion zones on Ti. The interaction between lattice vibrations and polarization, caused by out-of-phase atomic vibrations in ionic or polar crystals, explains the LO-TO splitting. This phenomenon occurs only in optical phonons, which arise when the cation Ti and anion Sn vibrate in opposite directions, creating an oscillating dipole. Long-range dipole–dipole interactions significantly impact the optical branches, causing the longitudinal and transverse optical modes to split more near the $\Gamma$-point. 

The phonon spectra obtained from both GGA-PBE and LDA-PZ are comparable, as seen in the phonon density of states (PDOS), where the LO, TO, LA, and TA branches are clearly distinguishable. The optical branches, particularly the longitudinal ones, remain flat due to the mass mismatch between Ti and Sn. As a result, the acoustic and optical branches are separated by a phonon gap appearing between 124 to 215 cm$^{-1}$ in LDA-PZ and 116 to 198 cm$^{-1}$ in GGA-PBE. Although experimental data for TiSn are unavailable, the computed optical phonon spectra closely resembles those reported for a similar zincblende compounds, such as ZnSe \cite{znse}. To further investigate the LO-TO splitting in TiSn, the screened effective charges were calculated using both LDA-PZ and GGA-PBE functionals:
\begin{equation}
    Z^* = \frac{Z_B}{\varepsilon(\infty)}
\end{equation}
In LDA-PZ, the Born effective charge of Ti is $Z_B = 4.56$ and the long wavelength dielectric constant is $\varepsilon(\infty) = 40.10$, yielding a screened effective charge ($Z^*$) of 0.721, while in GGA-PBE, with Born effective charge of Ti, $Z_B = 3.98$ and $\varepsilon(\infty) = 29.97$, resulting in $Z^* = 0.726$. The computed screened effective charges suggest that zb-TiSn behaves similar to the established zincblende semiconductors, such as ZnO ($Z^* = 0.923$), ZnS ($Z^* = 0.828$), ZnSe ($Z^* = 0.768$), and ZnTe ($Z^* = 0.704$)\cite{znse}. 

In zb-TiSn, the longitudinal optical (LO) phonon frequency is higher than the transverse optical (TO) phonon frequency at the X symmetry point. At this \textbf{q} point, the heavier atom (Sn) primarily participate in the LA phonon mode, while the lighter atom (Ti) vibrates with the LO phonon mode. The calculated frequency ratio is $\frac {\omega_{X}{LO}}{\omega_{X}LA} = 1.8$ for GGA-PBE and $1.9$ for LDA-PZ closely matches the square root mass ratio $\sqrt{m_{Sn} / m_{Ti}} = 1.56$, or approximately $1.6$. Minor discrepancies arise due to LDA-PZ tendency to overestimate and GGA-PBE’s tendency to underestimate phonon frequencies. 

According to Srivastava et al.,\cite{srivastava} the hypothesis that any zincblende (ZB) or rock-salt material with a mass ratio greater than two would satisfy the condition $\omega_{X}({LO}) > \omega_{X}({TO})$ holds true for TiSn, reinforcing its lattice dynamics consistent with other zincblende compounds.

The phonon density of states (PDOS) provides insight into the vibrational properties of the constituent atoms. The low-frequency region, ranging from approximately zero to  $130$ cm$^{-1}$ in LDA-PZ and zero to $128$ cm$^{-1}$ in GGA-PBE, corresponds to acoustic phonon modes dominated by Sn vibrations. Due to the mass difference, the lighter Ti atoms exhibits higher vibrational frequencies, leading to higher optical phonon frequencies. 

\section{Optical Properties}

Studying the optical properties of narrow band gap semiconductors is crucial for optimizing their performance for various applications, such as photo-detectors, solar cells, and infrared sensors. Understanding how these materials interact with light will enable the design of devices with improved efficiency and reduced energy losses.

\begin{figure}
    \centering
\subfigure[LDA-PZ]{\includegraphics[scale=0.3]{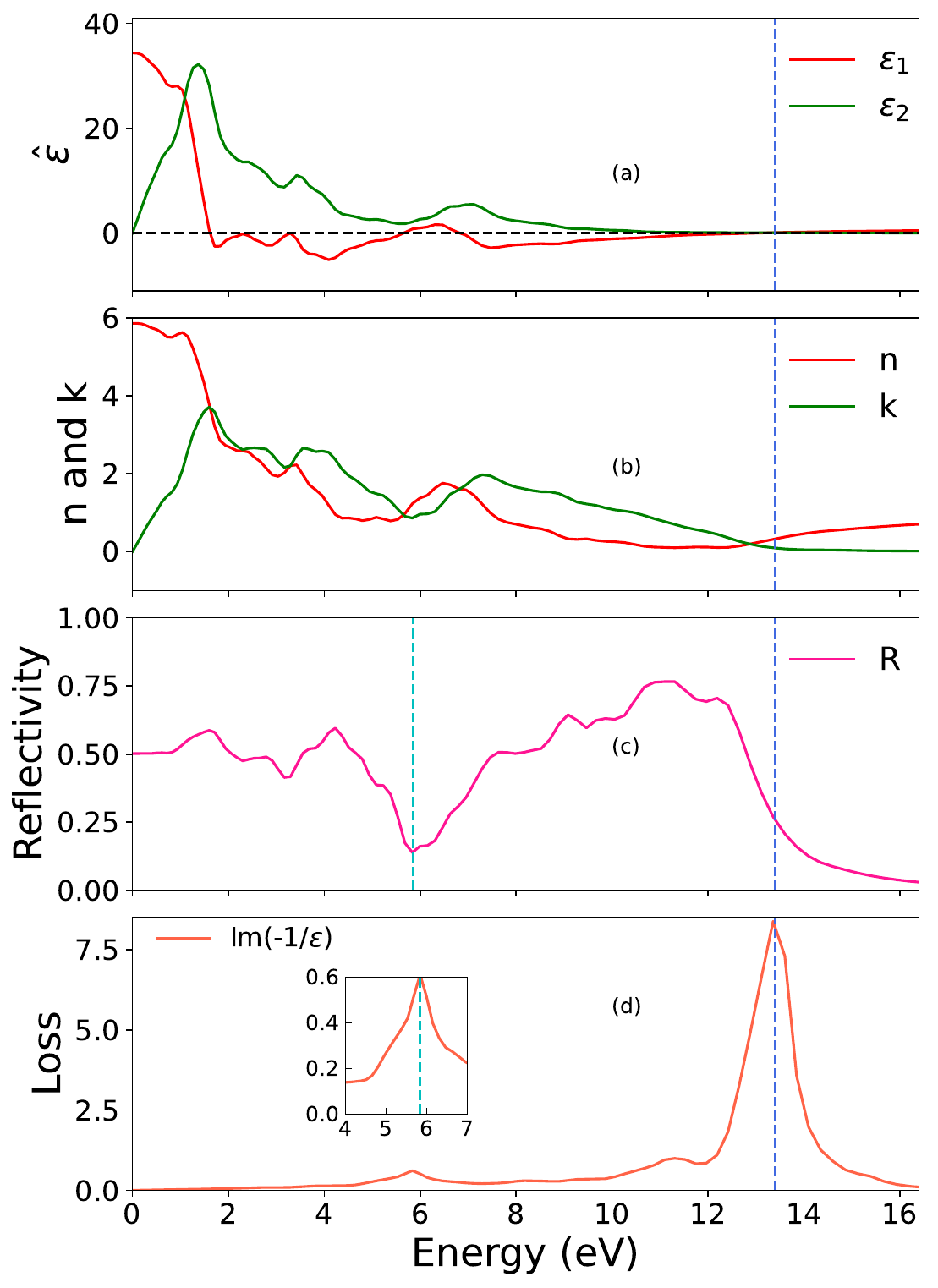}}
\subfigure[GGA-PBE]{\includegraphics[scale=0.3]{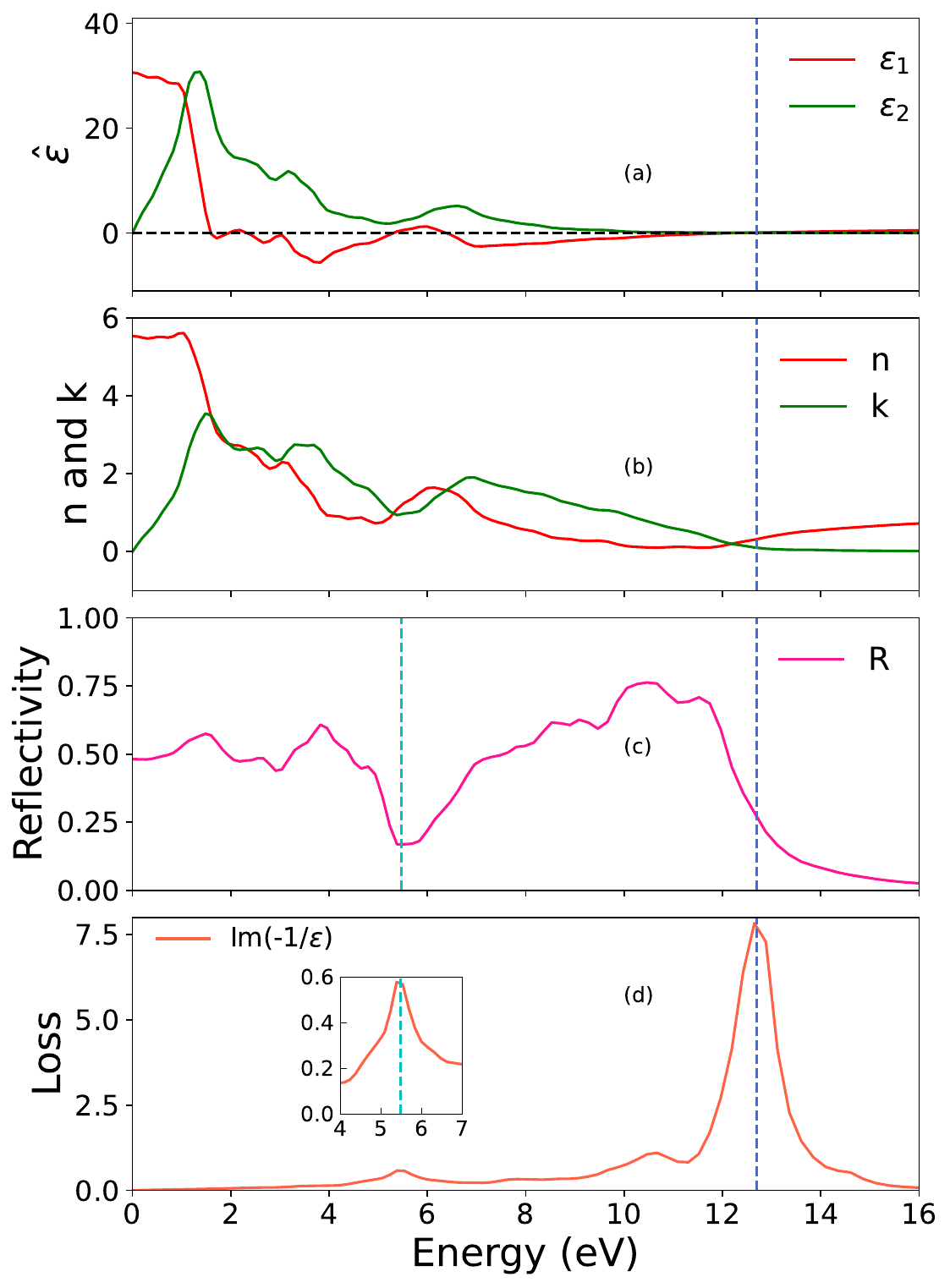}}
\subfigure[GGA-RPBE]{\includegraphics[scale=0.3]{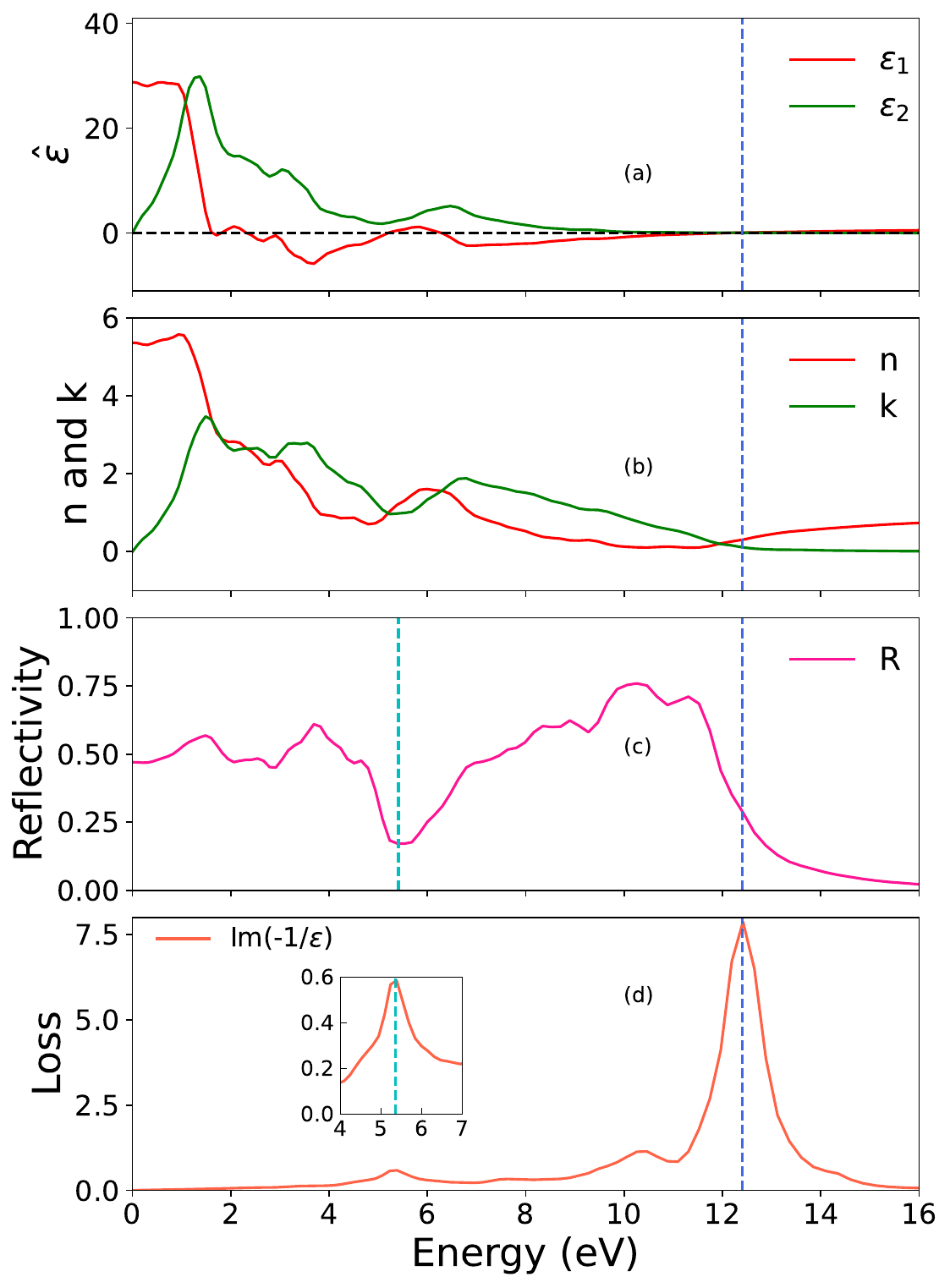}}
    \caption{Optical properties of TiSn under different approaches:  
    (i) LDA-PZ, (ii) GGA-PBE, and (iii) GGA-RPBE.  
    a) The real part $\varepsilon_1 (\omega)$ and  imaginary part $\varepsilon_2 (\omega)$,  
    b) Refractive index $n$ and extinction coefficient $k$,  
    c) Reflectivity $R(\omega)$ and  
    d) Energy loss spectrum $L(\omega)$ are shown.}
    \label{optical}
\end{figure}

We analyze the linear optical properties of zb-TiSn including the refractive index `n', extinction coefficient `k', reflectivity `R' of zb-TiSn. These optical properties of TiSn are studied through the dielectric function, which describes the linear response of the system to light. Within the Random Phase Approximation (RPA) , the dielectric function is derived from the  electronic band structure, calculated using LDA-PZ, GGA-PBE  and GGA-RPBE approximation as shown in Figure\ref{optical}a,b and c. However, we have extended out calculations using GLLB-sc functional also which could determine the band gap energy accurately compared to the functional mentioned before. The resulting band gap from GLLB-sc is 0.38 eV and it is very similar to the one obtained using GGA-PBE  and GGA-RPBE. Hence we have chosen not to present the optical properties from GLLB-sc in the manuscript, however we have provide them in the section-8 of the supplementary data file for readers convenience.

The dielectric function $\varepsilon (\omega) = \varepsilon_1 (\omega) + i \varepsilon_2 (\omega)$, describes the optical response of the medium as a function of frequency $\omega$. Where, the real part $\varepsilon_1 (\omega)$ of the dielectric function represents polarization response of the material to the incident light, describing the  the non-dissipative "screening" effect that influences the light propagation through the material. In contrast, the imaginary part $\varepsilon_2 (\omega)$ corresponds to the dissipative response of electrons to incoming electromagnetic waves, leading to energy loss as the wave propagates through the material. The expression for the imaginary part of the dielectric function $\varepsilon (\omega)$ is given by\cite{imaginary}

\begin{equation}
    \varepsilon_2(\omega) = \left( \frac{4\pi^2 e^2}{m^2 \omega^2} \right) \sum_{i,j}
    \int_{\mathbf{k}} \left| \langle i | M | j \rangle \right|^2 f_i (1 - f_j) \delta(E_{j,\mathbf{k}} - E_{i,\mathbf{k}} - \omega) d^3k
\end{equation}

where, $M$ is the dipole matrix, $i$ and $j$ represent the initial and final states respectively, $f_i$ the Fermi distribution function for the $i$-th state, and $E_i$ is the energy of electron in the $i$-th state with crystal wave vector $\mathbf{k}$.

By using the Kramers-Kronig dispersion relations\cite{wooten}\cite{kumk}, the real part $\varepsilon_1 (\omega)$ can be extracted:

\begin{equation}
	\varepsilon_1 (\omega) - 1 = \frac{2}{\pi} P \int_0^{\infty} \frac{\omega' \varepsilon_2 (\omega')}{(\omega')^2 - \omega^2} d\omega'
\end{equation}

\begin{equation}
	\varepsilon_2 (\omega) = -\frac{2\omega}{\pi} P \int_0^{\infty} \frac{\varepsilon_1 (\omega') - 1}{(\omega')^2 - \omega^2} d\omega'
\end{equation}
\\
where $P$ denotes the principal value of the integral.

The reflectivity $R$ of solids at normal incidence is given by Fresnel’s relation:

\begin{equation}
	R = \frac{(1 - n)^2 + k^2}{(1 + n)^2 + k^2}
\end{equation}
\\
where the refractive index $n$ and extinction coefficient $k$ are:
\\

\begin{equation}
	n = \sqrt{\frac{1}{2} \left[ \left( \varepsilon_1^2 + \varepsilon_2^2 \right)^{1/2} + \varepsilon_1 \right]}
\end{equation}

\begin{equation}
	k = \sqrt{\frac{1}{2} \left[ \left( \varepsilon_1^2 + \varepsilon_2^2 \right)^{1/2} - \varepsilon_1 \right]}
\end{equation}
\\
From Drude model the energy loss spectrum is given by,
\begin{equation}
	L(\omega) \propto Im \left( -\frac{1}{\varepsilon(E)} \right) = \frac{\varepsilon_{2}}{\varepsilon_1^{2} + \varepsilon_2^{2}}.
\end{equation}

As shown in figure \ref{optical}a, the real part of the dielectric function $\epsilon_{1}(\omega)$ exhibits normal dispersion until 1.08 eV after which it undergoes anomalous dispersion, indicating the small band gap and high refractive index of the material. The refractive index $n(0) \approx 5.86$ (LDA-PZ),5.52 (GGA-PBE) and 5.35 (GGA-RPBE) initially marks high polarizability as shown in figure \ref{optical}b. $\epsilon_{1}(\omega)$ crosses zero at 1.62 eV and remains below zero until 5.66 eV further supporting the small band gap nature of the zb-TiSn. The effect of GGA-PBE and GGA-RPBE is relatively less on the linear optical properties and they appear similar to LDA-PZ. The imaginary part of the dielectric function $\epsilon_{2}(\omega)$ reaches a peak maxima at 1.34 eV marking the inflection point in $\epsilon_{1}(\omega)$ as shown in figure \ref{optical}a. 

The reflectivity curve of the zb-TiSn semiconductor is shown in Figure \ref{optical}c. The reflectivity reaches an average of 50\% in the infrared and visible regions, with a maximum of 76\% at 11.06 eV(LDA-PZ),10.43 eV (GGA-PBE) and 10.26 eV(GGA-RPBE). The short valley at 5.87 eV (LDA-PZ),5.43 eV(GGA-PBE)and 5.36 eV (GGA-RPBE) corresponds to the surface plasmon frequency, while the peak at 16.31 eV(LDA-PZ),16.80(GGA-PBE) and 15.50 eV(GGA-RPBE) marks the volume plasmon frequency. The plasmon frequencies are distinctly visible in the loss spectra as shown in figure \ref{optical}d. By tailoring the optical characteristics of narrow band gap semiconductors, photon capture can be enhanced, optical losses in waveguides can be minimized and more efficient optoelectronic devices can be designed and developed, particularly for infrared and near-infrared applications.

\textcolor{blue}{Light absorption is a key factor in determining the properties like photovoltaic and photocatalytic efficiency\cite{omega}, we have calculated the absorption coefficient of zb-TiSn with PBE functional, the absorption coefficient $I(\omega)$  is given by \cite{omega1}}

\begin{equation}
I(\omega) = \frac{\omega \varepsilon_2}{c n}
\end{equation}

Where $c$ is light velocity, $n$ is the refraction index:

\begin{equation}
I(\omega) = \sqrt{2} \, \omega \left[ \sqrt{\varepsilon_1^2(\omega) + \varepsilon_2^2(\omega)} - \varepsilon_1(\omega) \right]^{1/2}
\end{equation}    

\begin{figure}[h]
\centering
\includegraphics[scale=0.65]{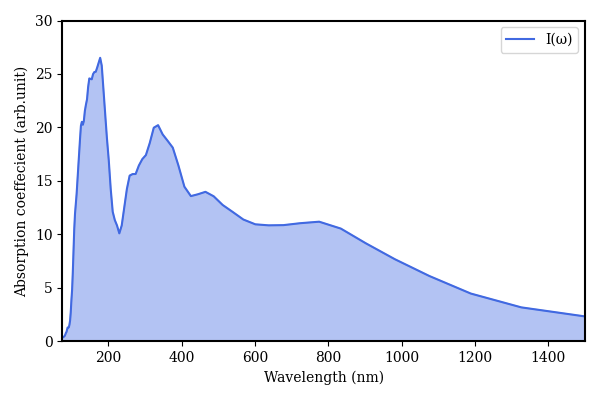}
\caption{Absorption spectrum of zb-TiSn with GGA-PBE functional}
\label{fig:absorption_spectrum}
\end{figure} 
                                                                               
 \textcolor{blue}{The absorption spectrum using GGA-PBE is shown in Figure~\ref{fig:absorption_spectrum}, marks effective absorption from 100 to 1500 nm, covering the ultraviolet (UV), visible (VIS), and near- infrared (NIR) regions.These results indicate that zb-TiSn is highly effective in absorbing radiation within this range. 
The transition of the generated free charge carriers are driven by the energy absorbed from the entire energy range of 100 to 1500 nm, supporting efficient optoelectronic performance, particularly in applications like low-energy photodetectors for infrared region and waste heat harvesting.(For the absorption spectra of other functionals, please refer section-9 in the supplementary data file )}

\section{Conclusion}

Through first-principles calculations based on density functional theory, we have investigated the zincblende structure of TiSn and identified it as a narrow bandgap semiconductor based on its electronic band structure. The structural properties were determined by optimizing the energy eigenvalues and fitting them to the Murnaghan equation of state to extract the lattice constant, equilibrium volume, bulk modulus, and its derivative for the ground state of zb-TiSn. The electronic band structure was computed along the high-symmetry path (L-$\Gamma$-X-K-$\Gamma$) of the Brillouin zone, and the carrier effective mass near the conduction band minimum (CBM) and valence band maximum (VBM) was determined. The charge density distribution plot indicates the presence of covalent bonding through charge sharing, while Bader charge analysis provides evidence of ionic bonding by measuring charge transfer. The optical properties of zb-TiSn were analyzed using the complex dielectric function and Kramers-Kronig relations, revealing its sensitivity to infrared (IR) radiation. All calculations were performed using three different exchange-correlation functionals, LDA-PZ, GGA-PBE, and GGA-RPBE to assess the robustness of the semiconductor material.

\section*{Conflicts of interest}
There are no conflicts to declare.

\section*{Acknowledgements}

\textbf{SR},\textbf{SM} and \textbf{UMKK} would like to thank \textbf{Mr.Mohan Kumar C M}(Deputy Director-CTS-VIT) for providing us with computational resources.  The authors acknowledge\textbf{ VIT SEEDGRANT-SG20240040}supporting the current research project.

\section*{Author contributions}
\textbf{Sudeep R } contributed to DFT calculations for electronic structure, Investigation, Visualization and first draft preparation; 
\textbf{Sarojini M} also equally contributed to DFT calculations for lattice dynamics, Investigation, Visualization and first draft preparation;
\textbf{Uma Mahendra Kumar K}: Conceptualization, Methodology, Visualization, Investigation and Writing- Reviewing and Editing.

\bibliographystyle{elsarticle-num}
\bibliography{tisn_ref}
\end{document}